# Shear softening of Earth's inner core indicated by its high Poisson's ratio and elastic anisotropy

Zhongqing Wu[1]

1. Laboratory of Seismology and Physics of Earth's Interior, School of Earth and Space Science, University of Science and Technology of China, Hefei, Anhui, 230026, PR China

2. Mengcheng National Geophysical Observatory, Anhui, PR China.

**Key Points:**

Shear softening causes an unusually large Poisson's ratio and a noticeable elastic anisotropy.

The inner-core may undergo shear softening.

The identification of the elements that can stabilize bcc iron provides critical information on the composition of the inner core.

**Earth's inner core exhibits an unusually high Poisson' ratio and noticeable elastic anisotropy. The mechanisms responsible for these features are critical for understanding the evolution of the Earth but remain unclear. This study indicates that once the correct formula for the shear modulus is used, shear softening can simultaneously explain the high Poisson's ratio and strong anisotropy of the inner core. Body-centred-cubic (bcc) iron shows shear**

[1] wuzq10@ustc.edu.cn

instability at the pressures found in the inner-core and can be dynamically stabilized by temperature and light elements. It is very likely that some combinations of light elements stabilize the bcc iron alloy under inner-core conditions. Such a bcc phase would exhibit significant shear softening and match the geophysical constraints of the inner core. Identifying which light elements and what concentrations of these elements stabilize the bcc phase will provide critical information on the light elements of the inner core.

I. Introduction

It has been well established by geochemical, geophysical, and mineral physical investigations that the Earth's core mainly consists of an iron-nickel alloy. The densities of the iron-nickel alloy at the pressure and temperature (*PT*) conditions at the Earth's core are slightly higher than the density of the Earth's core itself[*Birch*, 1952], which suggests that the inner core must contain small amounts of lower atomic weight elements (light elements). Several light elements, such as silicon, sulphur, oxygen, carbon, hydrogen, and magnesium, have been proposed to exist in the Earth's core based on their effects on the density and sound velocities of the iron alloy[*Hirose et al.*, 2013; *Kadas et al.*, 2009]. Although the light-element-bearing iron alloy could simultaneously produce the densities and P-wave velocities of Preliminary reference Earth model (PREM)[*Dziewonski and Anderson*, 1981], its S-wave velocities are generally much larger than those of PREM. The Poisson's ratio of an iron alloy with a hexagonal-close-packed (hcp) phase is approximately 0.36[*Lin et al.*, 2003; *Mao et al.*, 2001; *Murphy et al.*, 2013; *Shibazaki et al.*, 2012], which is much smaller than the 0.44 Poisson's ratio of the Earth's inner core. The low S-wave velocities and, conversely, the high Poisson's ratio, of the Earth' inner core are an enigma.

The main mechanism that has been proposed for the high Poisson's ratio of the inner core includes partial melting[*Vocadlo*, 2007], the premelting effect[*Martorell et al.*, 2013], and the existence of an iron-carbides inner core[*Chen et al.*, 2014; *Gao et al.*, 2008; *Prescher et al.*, 2015]. However, the melting of the top of the Earth's inner core does not easily explain the high Poisson's ratio throughout the entire inner

core[*Gubbins et al.*, 2011]. The premelting mechanism also requires T~0.99$T_m$ (where $T_m$ is the melting temperature) throughout the entire inner core. By extrapolating the measured results below 200 GPa to the inner-core conditions, $Fe_3C_7$ has Poisson's ratios similar to those of the Earth's inner core[*Chen et al.*, 2014; *Prescher et al.*, 2015]. However, first-principles calculations[*Raza et al.*, 2015] show that the Poisson's ratio changes slightly from 0.37 at 150 GPa to 0.38 at 360 GPa, which is different from the extrapolated experimental result.

Another important feature of the Earth's inner core is its elastic anisotropy. P-wave velocities are ~3-4% faster along the polar axis than in the equatorial plane[*Creager*, 1992; *Song*, 1997; *Tromp*, 1993; 2001]. The inner layer of the inner core has a stronger anisotropy than the outer layer[*Ishii and Dziewonski*, 2002; *Niu and Chen*, 2008; *Sun and Song*, 2008; *Wang et al.*, 2015]. The anisotropy can be caused by the lattice preferred orientation (LPO) of iron alloy developed under plastic flow conditions. Hcp iron shows a degree of elastic anisotropy at the inner-core density when temperature effects are excluded[*Stixrude and Cohen*, 1995]; however, it is probably nearly isotropic at high temperatures[*Belonoshko et al.*, 2003; *Mattesini et al.*, 2010; *Sha and Cohen*, 2010]. Although the partial melting and premelting effect of the hcp crystal may result in a high Poisson's ratio that matches that observed in the inner core, these explanations are confronted with challenges in interpreting the elastic anisotropy of the inner core. Whether pure $Fe_3C_7$ iron can produce the elastic anisotropy of the inner core remains unknown because of the absence of single-crystal elasticity of $Fe_3C_7$ with the Pbca space group[*Prescher et al.*, 2015] at the inner-core conditions. This study demonstrates that the high Poisson's ratio and strong anisotropy can be produced simultaneously by shear softening, which suggests that the inner core may be in a state of shear softening.

**II. Shear softening with high Poisson's ratio and strong anisotropy**

The cubic crystal has been chosen to allow for a concise discussion of the relationship between shear softening and the high Poisson's ratio. The main conclusions derived here can be applied to other types of crystals. The shear moduli of an aggregate have no exact solutions and their values are located between two bounds: an upper Voigt bound, which assumes that strain is uniform and that the stress is supported by the individual grains in parallel,

$$G_V = (2C' + 3C_{44})/5, \qquad (1)$$

and a lower Reuss bound, which assumes that the stress is uniform and the strain is the total sum of the strains of the individual grains,

$$G_R = \frac{5}{2/C' + 3/C_{44}}, \qquad (2)$$

where $C' = (C_{11} - C_{12})/2$. The Voigt-Reuss-Hill average (VRH), namely, the arithmetic average of the two bounds, is widely used for the shear moduli of an aggregate. Shear softening is assumed to occur in $C'$. The same conclusion can be obtained for shear softening of $C_{44}$. If a cubic crystal shows noticeable shear softening, the parameter $A = C'/C_{44}$ is near zero. Thus, the shear velocity along [110] direction is much smaller than the one along [100] direction. The crystal is highly anisotropic. The difference between the Voigt and Reuss bounds is sensitivity to $A$. If the crystal is isotropic, $A=1$ and $G_V=G_R=G_{VRH}$. Hcp iron at the inner-core conditions has been found to have small anisotropy, i.e., their $G_V \sim G_{VRH}$. Therefore, Eq. (1) can be used to calculate the shear modulus of hcp iron[*Vocadlo*, 2007]. However, the differences between the two bounds become particularly large for crystals that undergo noticeable shear softening. It is not suitable to adopt the Voigt bound as the aggregate shear modulus for anisotropic crystals such as bcc iron because the Voigt bound significantly overestimates the shear modulus. As shown in Fig. 1, the Voigt bound is much larger than the upper bound of the tighter Hashin-Shtrikman bounds[*Hashin and Shtrikman*, 1962]. Although the Voigt and Reuss bounds are significantly different from the upper and lower bounds of the Hashin-Shtrikman bounds, respectively, the VRH is almost identical to the arithmetic average of the Hashin-Shtrikman bounds[*Karki et al.*, 2001], which suggests that $G_{VRH}$ should be a good choice for the shear modulus of the aggregate and explains the popularity of the VRH method.

The shear softening of a crystal not only leads to its strong anisotropy but also its extremely high Poisson's ratio. The dependence of the Poisson's ratio $\nu$ on $A$ is shown in Fig. 2a, with fixed $C_{44}$ and bulk modulus values. For the isotropic crystal with $A=1$, the Poisson's ratio is chosen to be 0.38, which is similar to that of hcp iron[*Lin et al.*, 2003]. It is clear that shear softening noticeably increases $\nu$ even when $G_V$ is used instead of $G_{VRH}$. However, the largest $\nu$ at $A=0$ is 0.42, which is smaller than the $\nu$ of the inner core (~0.44). As shown in Fig. 2, $G_{VRH}$ is dramatically smaller than $G_V$, especially when $A$ is nearly 0. The largest $\nu$ calculated using $G_{VRH}$ was 0.46, which is larger than the inner-core value. The iron phase can reach the inner-core value of 0.44 when $A \sim 0.15$. The iron phase with the inner-core Poisson's ratio exhibits highly elastic anisotropy (Fig. 2b). Developing the LPO for a small portion of the iron alloy in the inner core is sufficient to explain the elastic anisotropy of the inner core. Therefore, shear softening can cause both a high Poisson's ratio and elastic anisotropy. The inner core may undergo shear softening.

**III. The possible phase with the shear softening**

It is worth investigating the shear softening of iron-alloy phases under inner-core conditions, especially for new, recently identified phases. Currently, we only know that the bcc phase experiences shear softening at high pressure as its shear modulus $C'$ decreases with increasing pressure. Similar shear softening occurs in stishovite and leads to the extremely large S-wave anisotropy and strong P-wave anisotropy of stishovite[*Karki et al.*, 1997; *Yang and Wu*, 2014]. Although the bcc phase shows shear instability at the pressures in the Earth's core[*Stixrude et al.*, 1994{Soderlind, 1996 #37}], the phase has been shown to be dynamically stable under inner-core conditions by numerous works, such as the classical[*Belonoshko et al.*, 2003; *Belonoshko et al.*, 2007; *Belonoshko et al.*, 2008] and ab initio[*Belonoshko et al.*, 2007; *Bouchet et al.*, 2013; *Cui et al.*, 2013; *Kadas et al.*, 2009; *Vocadlo*, 2007; *Vocadlo et al.*, 2003] molecular dynamic simulation and lattice dynamic calculations [*Luo et al.*, 2010]. Because the bcc phase has larger vibrational entropy than the hcp phase and light elements stabilize the bcc phase[*Belonoshko et al.*, 2003; *Vocadlo et*

*al.*, 2003], the bcc may be stable under inner-core conditions. The controversial results on the dynamic stability of the bcc phase under inner-core conditions from the ab initio calculations by Vocadlo et al[*Vocadlo et al.*, 2003] and by Godwal et al[*Godwal et al.*, 2015], with slightly different convergence-control parameters, imply that the bcc phase may only be stable at a particularly narrow temperature range just below the melting curve and may show noticeable shear softening. Several transition metals are known to have a stable bcc phase at a narrow temperature range below the melting curve[*Petry*, 1995]. Noticeable shear softening can lead to both a high Poisson's ratio and anisotropy.

Experimental results on the stable phase of the inner core are also controversial. Dubrovinsky et al.[2007] observed a bcc Fe-Ni alloy at 225 GPa and 3400 K. However, this phase transformation has not been reproduced by the experiments conducted at similar *PT* conditions with a similar alloy composition[*Sakai et al.*, 2011; *Sakai et al.*, 2012]. The reason for the divergence between the experiments remains unclear. The hysteresis effect is a possible factor. The bcc phase may only be stable at particularly narrow temperature ranges below its melting temperature. It is possible that iron retains its hcp phase even though the hcp phase is metastable under inner-core conditions. The hysteresis behaviour has been found in many first-order phase transformations such as supercooling and superheating. It is worth investigating the hysteresis behaviour of the first-order phase transformation of iron. Although the hcp phase for Fe and Fe-Si alloy has been observed under inner-core conditions[*Tateno et al.*, 2012; *Tateno et al.*, 2010; *Tateno et al.*, 2015], it is possible that particular combinations of light elements make the bcc phase stable under inner-core conditions due to the small difference in free energy between bcc and hcp pure iron under inner-core conditions[*Belonoshko et al.*, 2003; *Bouchet et al.*, 2013; *Dubrovinsky et al.*, 2007; *Kadas et al.*, 2009; *Komabayashi and Fei*, 2010; *Vocadlo et al.*, 2003]. Identifying the elements and the concentrations of elements that can stabilize the bcc phase will provide critical insight regarding the light elements in the inner core.

The elastic properties of the bcc iron phase match the high Poisson's ratio and anisotropy of the inner core. As shown in Table 1, the bcc phase reported by Belonoshko et al.[2007] is highly anisotropic, with $A$=0.155. These authors obtained the shear modulus using the Voigt bound, which is ~18% larger than the higher bound of the Hashin-Shtrikman bounds and is ~36% larger than $G_{VRH}$. The S-wave and P-wave velocities of the bcc phase under inner-core conditions match those of PREM well once $G_V$ is replaced with $G_{VRH}$. Correspondingly, the Poisson's ratio also increases from ~0.42 to ~0.44. Therefore, contrary to the claims of Belonoshko et al.[2007], the reported elasticity for the bcc phase suggests the origin of the low rigidity of the Earth's inner core. Belonoshko et al.'s result[2007] indicating that defects and grain boundaries reduce $G$ may also result from their calculation method. It is possible that the $G$ value with defects and grain boundaries calculated by Belonoshko et al.[2007] may not correspond to $G_V$ but to a value between $G_V$ and $G_R$. Thus, the extent that defects and grain boundary reduce $G$ using Belonoshko et al.'s method[2007] will depend dramatically on the anisotropy of the crystal. This result can be checked by performing a similar simulation for an elastic isotropic iron phase. Extensive research studies have shown that the measured $G$ for the aggregate agrees well with the calculated $G_{VRH}$ [Wu et al., 2013; Wu and Wentzcovitch, 2011], which reflects to some extent that the intrinsic effects of defects and grain boundaries on elasticity should be small.

Kadas et al.,[2009] found that $C'$ =6.4 GPa at 355 GPa and 5000 K for $Fe_{0.91}Mg_{0.09}$. Less Mg is required for dynamical stability at higher temperature. As shown in Fig. 2, their results can also be attributed to the effect of shear softening on the Poisson's ratio. We expect that the bcc phase with other light elements can also match the high Poisson's ratio of the Earth's inner core once the light elements can stabilize the bcc phase because the shear modulus is especially sensitive to $A$ (Fig. 2) and hence the concentrations of light elements under these circumstances. The Poisson's ratio of PREM can be matched by fine-tuning the concentrations of light elements. This feature will be particularly helpful in constraining the light elements at the inner core.

## IV. Conclusion

$G_V$ is much larger than $G_{HS+}$ for a strong anisotropic crystal (Fig. 2 and Table 1), which shows that the Voigt bound significantly overestimates the shear modulus of a strong anisotropic crystal. If $G_V$ is replaced with $G_{VRH}$, an iron alloy phase with a sufficient shear softening can match both the strong anisotropy and the high Poisson' ratio of the inner core. Bcc phase is a currently known iron phase undergoing shear softening at high pressure. The energy difference between bcc and hcp iron phase is small under the inner-core conditions. It is very likely that particular combinations of light elements can stabilize the bcc iron at the inner-core conditions.

The controversial results on the dynamic stability of bcc phase under the inner-core conditions by the same calculation method with slightly different convergence-control parameters [*Belonoshko et al.*, 2003] [*Bouchet et al.*, 2013] [*Godwal et al.*, 2015] [*Vocadlo et al.*, 2003] suggest (1) that bcc phase may only be stable at a particularly narrow temperature range just below the melting curve, which challenges the experimental observation of the bcc phase under the inner-core conditions and (2) bcc phase under the inner-core conditions should be near the dynamic stable and unstable boundary and should undergo dramatic shear softening, which leads to both a high Poisson's ratio and anisotropy. The *S*-wave velocity of the inner core with shear softening is highly sensitive to temperature and the concentrations of the light elements, which is particularly helpful in constraining the concentrations of the light elements in the inner core. Identifying the elements and the concentrations of elements that can stabilize the bcc phase will provide critical insight regarding the light elements in the inner core.

## Acknowledgments

This study is supported by State Key Development Program of Basic Research of China (2014CB845905), the "Strategic Priority Research Program" of the Chinese

Table 1 Reanalysing the elastic properties of bcc iron reported by Belonoshko et al.[2007]. In addition to the Voigt bound ($G_V$) of the shear modulus reported by Belonoshko et al. [2007], the Reuss bound ($G_R$), the Voigt-Reuss-Hill average ($G_{VRH}$), the upper Hashin-Shtrikman bound ($G_{HS+}$), the lower Hashin-Shtrikman bound ($G_{HS-}$), and the arithmetic average of the upper and lower Hashin-Shtrikman bound ($G_{HS}$) are listed. The S-wave velocities $V_S^V$ and $V_S^{VRH}$ are calculated using $G_V$ and $G_{VRH}$, respectively. $C' = (C_{11} - C_{12})/2$

| Parameter (units) | Ab initio | EAM | | Earth's IC |
|---|---|---|---|---|
| $P$ (GPa) | 356.7 | 360.0 | 360.0 | 363.9 |
| $T$ (K) | 6000.0 | 6000.0 | 7400.0 | 5000 to 8000 (5) |
| $B$ (GPa) | 1486.0 | 1372.7 | 1380.0 | 1425.3 |
| $C_{11}$ (GPa) | 1561.6 | 1391.0 | 1415.1 | |
| $C_{12}$ (GPa) | 1448. | 1363.5 | 1362.5 | |
| $C_{44}$ (GPa) | 365.5 | 448.0 | 387.0 | |
| $C'$ (GPa) | 56.8 | 13.8 | 26.3 | |
| $G_V$ (GPa) | 242.0 | 274.3 | 242.7 | |
| $G_R$ (GPa) | 115.1 | 32.9 | 59.7 | |
| $G_{VRH}$ (GPa) | 178.5 | 153.6 | 151.2 | 176.1 |
| $\rho$ (g/cm$^3$) | 13.58 | 13.90 | 13.78 | 13.09 |
| $V_L$ (km/sec) | 11.27 | 10.65 | 10.71 | 11.26 |
| $V_S^V$ (km/sec) | 4.22 | 4.44 | 4.20 | |
| $V_S^{VRH}$ (km/sec) | 3.63 | 3.32 | 3.31 | 3.67 |
| $G_{HS+}$ (GPa) | 205.6 | 210.7 | 193.6 | |
| $G_{HS-}$ (GPa) | 154.7 | 56.6 | 93.5 | |
| $G_{HS}$ (GPa) | 180.2 | 133.7 | 143.6 | |

**Figure Captions**

**Figure 1.** The shear modulus of bcc iron. $G_V$ and $G_R$ are the Voigt and Reuss bounds of the shear modulus. $G_{VRH}$ is the arithmetic average of $G_V$ and $G_R$. $G_{HS+}$ and $G_{HS-}$ are the upper and lower Hashin-Shtrikman bounds and $G_{HS}$ is the arithmetic average of $G_{HS+}$ and $G_{HS-}$. $A=(C_{11}-C_{12})/(2C_{44})$.

**Figure 2.** The dependence of Poisson's ratio (a), $G_{VRH}/G_V$ and single-crystal azimuthal anisotropy for P-wave $A_P = (V_P^{max} - V_P^{min})/V_P$ (b) on the $A=(C_{11}-C_{12})/(2C_{44})$, where $G_{VRH}$ and $G_V$ are the VRH average and Voigt bound of the shear modulus of the aggregate. A Poisson's ratio of 0.38 was chosen for $A=1$, which is close to the ratio for hcp iron. The bulk modulus and $C_{44}$ are fixed and $C'=1/2(C_{11}-C_{12})=AC_{44}$. Solid circles show Poisson' ratio based on the ab inito elastic modulus from Belonoshko et al. [2007], with the red circle calculated using $G_V$ and the green circle calculated using $G_{VRH}$. The solid star is based on the elastic modulus of bcc $Fe_{0.91}Mg_{0.09}$ from Kádas et al. [2009].

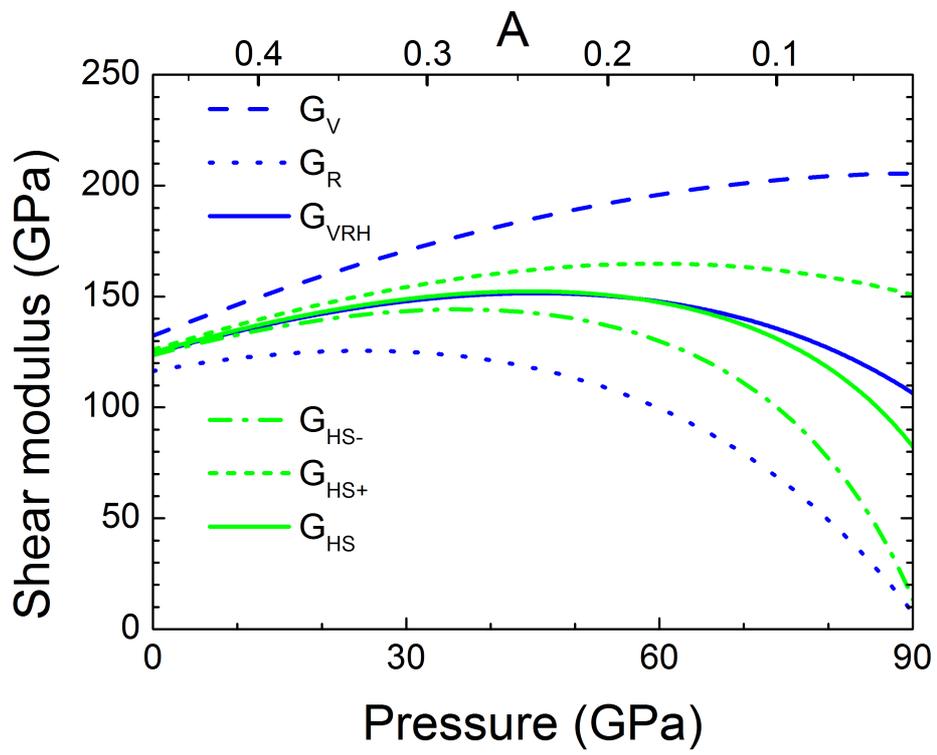

Figure 1

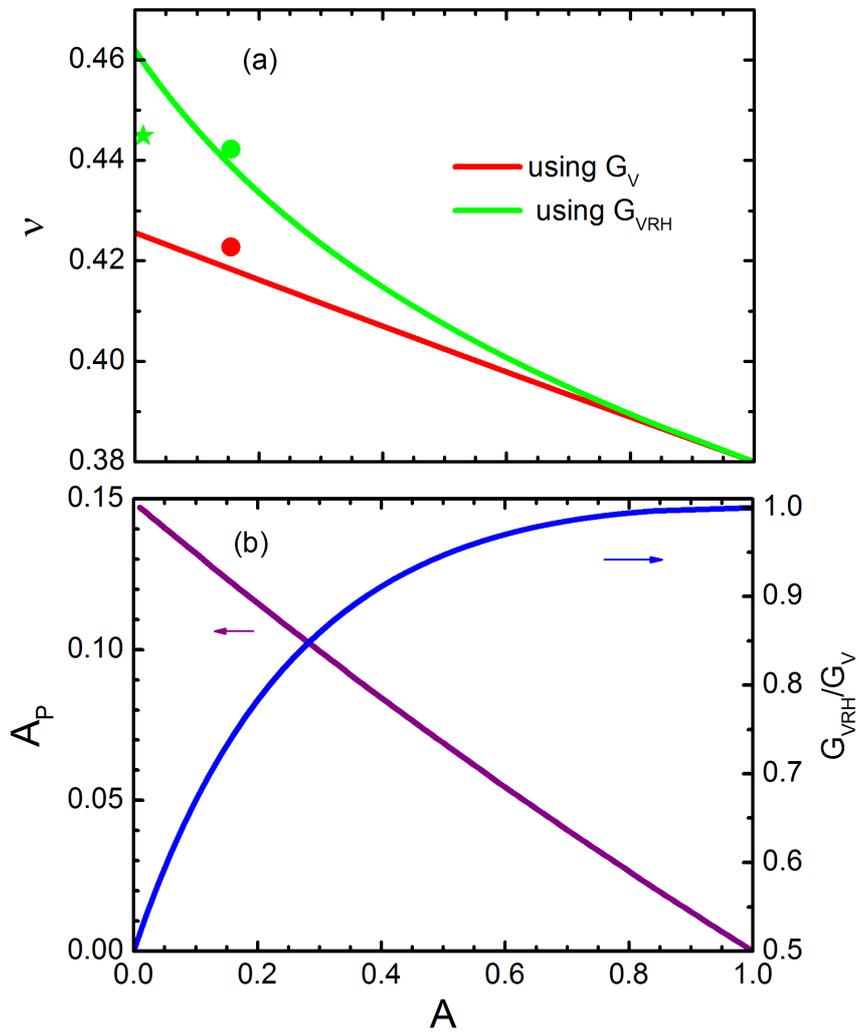

Figure 2